\documentclass{article}
\usepackage{physics}
\usepackage[labelfont=bf]{caption} 
\usepackage[font={small}]{caption}   
\usepackage[T1]{fontenc}
\usepackage{graphicx}
\usepackage{dcolumn}
\usepackage{bm}
\usepackage[pdftex,colorlinks=true,linkcolor=blue,citecolor=blue]{hyperref}
\usepackage[utf8]{inputenc}
\usepackage{lipsum}  
\usepackage{amsmath}
\usepackage{authblk}
\usepackage{amsfonts}
\usepackage{placeins}
\usepackage[numbers, sort&compress]{natbib} 
\usepackage{amssymb}
\usepackage{braket}
\usepackage{pdflscape}

    \usepackage[table, usenames, dvipsnames]{xcolor}
    \usepackage{makecell}
    \usepackage{boldline}
    \setcellgapes{3pt}
\usepackage{amsmath}
\usepackage{amssymb}

\usepackage{textgreek}

    \usepackage{lineno}  
    \usepackage{amsmath} 
    \usepackage{etoolbox} 

    \newcommand*\linenomathpatch[1]{%
    \cspreto{#1}{\linenomath}%
    \cspreto{#1*}{\linenomath}%
    \csappto{end#1}{\endlinenomath}%
    \csappto{end#1*}{\endlinenomath}%
    }

    \linenomathpatch{equation}
    \linenomathpatch{gather}
    \linenomathpatch{multline}
    \linenomathpatch{align}
    \linenomathpatch{alignat}
    \linenomathpatch{flalign}

\usepackage[strict]{changepage}
\usepackage{nicefrac}

    \newcommand{\C}{\hat{c}}

\usepackage{xspace}

\title{A mono-atomic orbital-based\\ 1D topological crystalline insulator}

\author{
    Gengming Liu $^{1\dag}$, Violet Workman $^{1\dag}$, Jiho Noh$^{2,3\dag}$, Yuhao Ma$^{1}$, \mbox{Taylor L. Hughes$^{1*}$,} \mbox{Wladimir A. Benalcazar$^4$,} Gaurav Bahl$^{2*}$ \\
    $^1$ Department of Physics, $^2$ Department of Mechanical Science $\&$ Engineering, \\
    University of Illinois at Urbana–Champaign, Urbana, IL 61801 USA \\
    $^3$ Sandia National Laboratories, Albuquerque, NM 87185, USA. \\
    $^4$ Department of Physics, Emory University, Atlanta, GA 30322, USA \\
}

\date{}  

\newcommand\blfootnote[1]{%
  \begingroup
  \renewcommand\thefootnote{}\footnote{#1}%
  \addtocounter{footnote}{-1}%
  \endgroup
}

\begin{document}  
\maketitle

\blfootnote{$^\dag$ These authors contributed equally. \qquad \qquad $^*$bahl@illinois.edu, hughest@illinois.edu}

\begin{abstract}
Topological crystalline insulators (TCIs) are classified by topological invariants defined with respect to the crystalline symmetries of their gapped bulk. The bulk-boundary correspondence then links the topological properties of the bulk to robust observables on the edges, e.g., the existence of robust edge modes or fractional charge. In one dimension, TCIs protected by reflection symmetry have been realized in a variety of systems where each unit cell has spatially distributed degrees of freedom (SDoF). However, these realizations of TCIs face practical challenges stemming from the sensitivity of the resulting edge modes to variations in edge termination and to the local breaking of the protective spatial symmetries by inhomogeneity. Here we demonstrate topologically protected edge states in a mono-atomic, orbital-based TCI that mitigates both of these issues. By collapsing all SDoF within the unit cell to a singular point in space, we eliminate the ambiguity in unit cell definition and hence remove a prominent source of boundary termination variability. The topological observables are also more tolerant to disorder in the orbital energies. To validate this concept, we experimentally realize a lattice of mechanical resonators where each resonator acts as an ``atom'' that harbors two key orbital degrees of freedom having opposite reflection parity. Our measurements of this system provide direct visualization of the $sp$-hybridization between orbital modes that leads to a non-trivial band inversion in the bulk. Furthermore, as the spatial width of the resonators is tuned, one can drive a transition between a topological and trivial phase. In the future we expect our approach can be extended to realize orbital-based obstructed atomic insulators and TCIs in higher dimensions. 
    
\end{abstract}

\maketitle

Topological crystalline insulators (TCIs) and obstructed atomic insulators  exhibit topological properties that are protected by the crystalline symmetries of their gapped bulk, e.g., quantized electric moments \cite{zak_berrys_1989, vanderbilt_electric_1993, king-smith_theory_1993, hughes_inversion-symmetric_2011, turner_quantized_2012, fu_topological_2011, benalcazar_electric_2017}. A guiding principle in topological matter is the bulk-boundary correspondence that establishes connections between bulk properties and boundary phenomena, e.g., a quantized bulk polarization or quadrupole moment can manifest as boundary-localized fractional charges and in-gap boundary modes \cite{Resta_macroscopic_1994,su_solitons_1979,guido_higher_2018,benalcazar_electric_2017,Benalcazar_Quantized_2017,Benalcazar_Quantization_2019,peterson_fractional_2020, peterson_quantized_2018, peterson_trapped_2021, yamada_bound_2022}. Notably, these boundary charges and states are robust against disorder as long as the bulk bandgap and the protective crystalline symmetries are preserved, and these, as well as other TCI phenomena, have spurred significant research interest because of their relevance to both science and engineering \cite{qi_topological_2011, hasan_colloquium_2010, mellnik_spin-transfer_2014, mong_universal_2014}.

\vspace{4pt}

To generate TCI band topology in a lattice system one needs more than one degree of freedom per unit cell. Most experimental realizations of TCIs have leveraged spatially distributed degrees of freedom (SDoF) within each unit cell and are susceptible to two important practical issues. The first originates from the ambiguity inherent in defining the unit cell itself. In particular, the bulk-boundary correspondence of obstructed atomic insulators can depend on the choice of unit cell \cite{lim_geometry_2015, bena_remarks_2009, fuchs_topological_2010, fruchart_parallel_2014, cayssol_topological_2021}. A canonical example of an obstructed atomic insulator, which we term as a 1D TCI for brevity in this manuscript, is the Su-Schrieffer-Heeger (SSH) chain composed of two spatially distinct sites within each unit cell \cite{su_solitons_1979}; its topology can be flipped between trivial and non-trivial by a mere translation of half a lattice constant in the assignation of the unit cell \cite{fuchs_orbital_2021}. Correspondingly, this implies that the topological boundary states of the SSH chain can be easily removed by purely surface effects, e.g., altering the last site in the chain with a vacancy or local potential. The second issue arises from the non-ideal nature of all experimental realizations--spatial disorder internal to a unit cell can result in local breaking of the protective crystalline symmetries, leading to distortions of the bandstructure and loss of topological protection.

\begin{figure}[p]
    \begin{adjustwidth*}{-1in}{-1in}
        \includegraphics[width=1.4\textwidth]{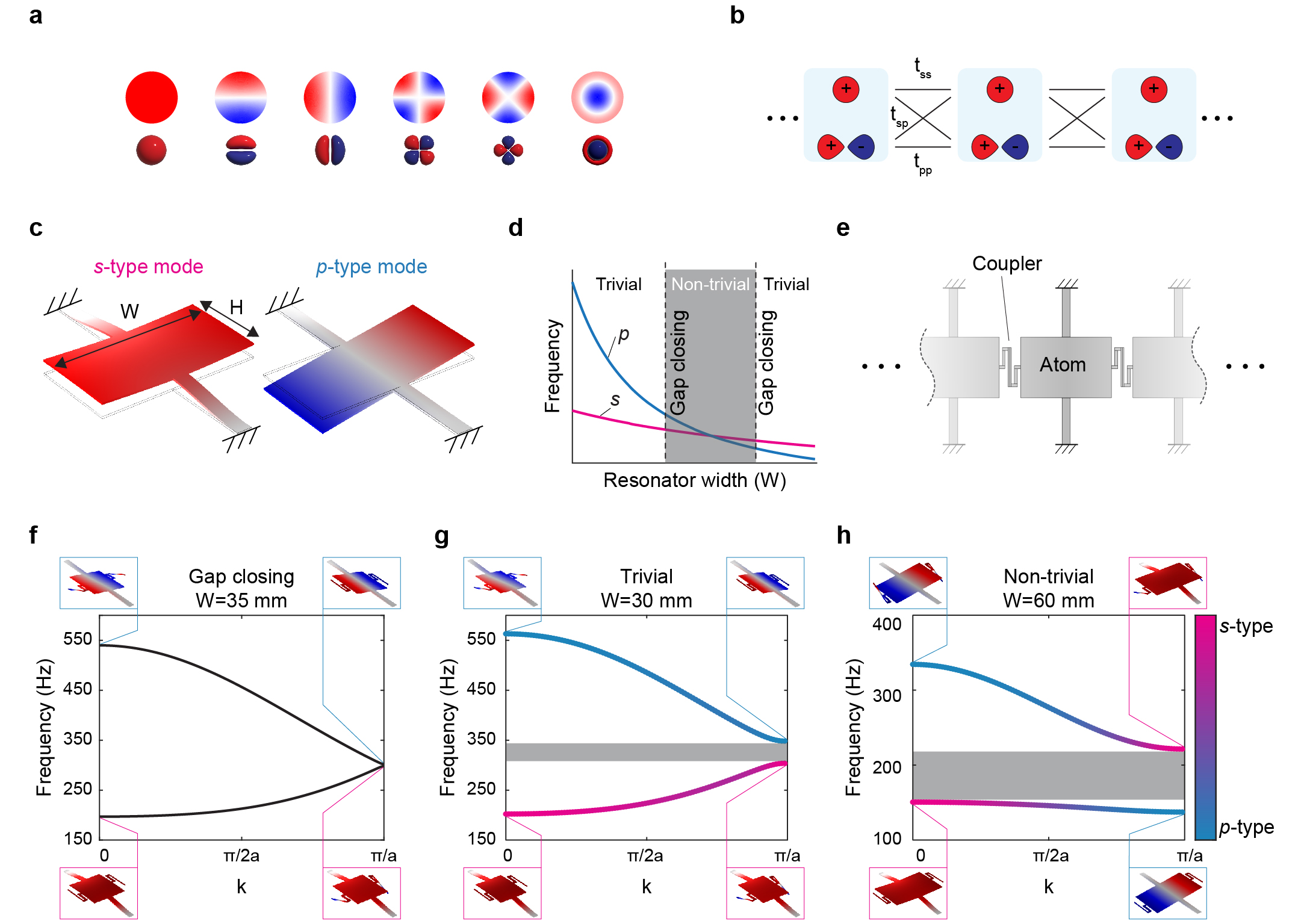}
    \centering
    \caption{
        \textbf{Principle and design for a one dimensional (1D) mono-atomic orbital-based insulator.} 
        \textbf{(a)} Many vibrational modes of
        a circular plate (above) are equivalent to atomic orbital modes (below).
        \textbf{(b)} The simplest 1D mono-atomic insulator composed of one $s$ and one $p$ orbital on each atom. The positive inter-atomic hopping strengths are indicated as ${t_{ss}}$, ${t_{sp}}$ and ${t_{pp}}$.
        \textbf{(c)} Our proposed implementation using a rectangular plate anchored with a central tether, has two orthogonal vibrational modes that are equivalent to $s$ and $p$ orbitals. 
        \textbf{(d)} The frequencies of the orbital modes can be tuned by varying the plate dimensions. Here we show tuning through the width parameter $W$. Trivial and non-trivial regions can be identified through the inequality presented in Eq.~{\ref{topological criterion}} and the dashed lines represent the gap closing points.
        \textbf{(e)} Bonding between adjacent sites can be achieved by means of thin coupling beams. The beams are wrapped into serpentine geometry for space efficiency.
        \textbf{(f)} Bandstructure simulation with periodic boundary condition (matching the structure shown in (d)) exhibits no bandgap for $W=35$ mm. Here, $a$ is the lattice constant, and the height parameter is fixed at $H = 40$ mm. 
        \textbf{(g)} Setting $W=30$ mm opens a trivial bandgap.
        \textbf{(h)} Setting $W=60$ mm produces a non-trivial bandgap. Mode shapes at high symmetry points are shown in insets. The $s$/$p$ characteristics of the modes are represented by band color (see Methods for projection calculation).
    }
    \label{concept}
    \end{adjustwidth*}
\end{figure}
\vspace{4pt}

Both these issues can be mitigated by collapsing all the SDoF within a unit cell to a single point. Indeed, one can replace the spatially-distributed basis of the unit cell with an orbital basis of a single ``atom'' \cite{fuchs_orbital_2021,minguzzi_topological_2022,McCann_catalog_2023}. The spatial symmetries are then determined by the symmetry representations obeyed by each independent orbital degree of freedom (ODoF) \cite{bradlyn_topological_2017}. In this context, hybridization between orbitals in neighboring atoms can drive a phase transition from a trivial atomic insulator state to a covalent obstructed atomic insulator corresponding to the formation of non-trivial bulk polarization and the appearance of protected in-gap edge states. Since there are no spatially separated modes within the unit cell, the ambiguity in how DoFs are assigned to each unit cell is removed, i.e., each unit cell holds one atom. Furthermore, there is no longer an ambiguity in the bulk-boundary correspondence since partial spatial cuts that might split the DoF in the boundary unit cell are not possible, and edge potentials affect each of the ODoF in the similar way. Moreover, the topology is more tolerant to disorder of the onsite mode energies since each site retains its exact representations (orbitals) of the reflection symmetry, unlike a comparable SDoF system (see Supplement \S\ref{sec:sup_robustness_to_onsite}).

\vspace{4pt}
While there is clear merit for an mono-atomic orbital-based TCI, to date no such TCI with such robust edge states has been realized. The direct manipulation of the degeneracy and spatial distribution of atomic orbitals in natural crystals can be extremely challenging. As a result, past experimental investigations of orbital physics have leveraged more flexible synthetic materials produced through a variety of approaches, such as with ultracold atoms trapped in optical lattices \cite{bloch_many-body_2008,bloch_ultracold_2005,li_physics_2016,wirth_evidence_2011,soltan-panahi_quantum_2012,slot_experimental_2017,jaksch_cold_1998,minguzzi_topological_2022}, polariton lattices \cite{milicevic_type-iii_2019,milicevic_orbital_2017,jacqmin_direct_2014,st-jean_lasing_2017}, photonic waveguide arrays \cite{jorg_artificial_2020,schulz_photonic_2022,zhang_realization_2023,guzman-silva_experimental_2021,caceres-aravena_Controlled_2022}, and nanomechanical resonators \cite{ma_nanomechanical_2021}. For instance, one can readily find a wealth of vibrational modes in thin plates (Fig.~\ref{concept}a) that have symmetries resembling $s$, $p$, and $d$ atomic orbitals, allowing the plate to be treated as an artificial atom. In this study, we distill this concept to realize the first orbital-based mono-atomic TCI having robust ODoF at the unit cell level, and a clear bulk-boundary correspondence that yields edge states independent of the number of atomic sites. We also experimentally observe the hybridization between the orbital states that opens the non-trivial, inverted bandgap.
   
\vspace{4pt}

To explore this concept, we choose a simple one dimensional (1D) mono-atomic insulator composed of one $s$ and one $p$ orbital on each atom. More precisely, we are considering an obstructed atomic insulator phase protected by reflection symmetry, and our ODoF consist of one even-parity ($s$) and one odd-parity ($p$) symmetry representation. Theoretical predictions and the topological properties of similar models have been presented in previous works \cite{shockley_surface_1939, vanderbilt_electric_1993, fuchs_orbital_2021, minguzzi_topological_2022,bradlyn_topological_2017,caceres-aravena_topological_2020}, and a tight binding model describing this material is illustrated in Fig.~\ref{concept}(b). Hopping strengths $t_{ss}$, $t_{pp}$, and $t_{sp}$ are indicated pairwise between $s$--$s$, $p$--$p$, and $s$--$p$ orbitals on adjacent atoms, respectively. Reflection symmetry prevents any inter-orbital interactions on the same atomic site. We fix all hopping strengths to be positive, and the momentum ($k$)-space Bloch Hamiltonian in the orbital basis is
\begin{equation}\label{Hamiltonian}
H(k)=
        \begin{bmatrix}
            E_s+2t_{ss}\cos{ka}&&2it_{sp}\sin{ka}\\
            -2it_{sp}\sin{ka}&&E_p-2t_{pp}\cos{ka}\\
        \end{bmatrix},
\end{equation}
where $a$ is the lattice constant, and $E_s$ ($E_p$) is the energy of the $s$ ($p$) orbitals. The relative signs of terms in $H(k)$ account for the parities of the orbital wavefunctions and would not change if the $s$ and $p$ orbitals are replaced by other orbitals with the same parity representations. 

To gain some intuition about this Bloch Hamiltonian we can expand {Eq.~(\ref{Hamiltonian})} in terms of Pauli matrices $\hat\sigma_i$ to find
\begin{align}
    H(k) = & \left[(t_{ss}-t_{pp})\cos{ka}+\frac{E_s+E_p}{2}\right]\mathbb{I} \notag \\
      & -(2t_{sp}\sin{ka})\hat{\sigma}_y+\left[(t_{ss}+t_{pp})\cos{ka}+\frac{E_s-E_p}{2}\right]\hat\sigma_z,
\label{Hamiltonian in Pauli matrices}
\end{align}
where $\mathbb{I}$ is the $2 \times 2$ identity matrix. The term proportional to the identity matrix affects the band dispersions, but not the band topology. In fact, the topology is determined by the signs of the $\hat{\sigma}_y$ terms at $k=0, k=\pi/a$\cite{hughes_inversion-symmetric_2011,turner_quantized_2012}. Indeed, for this model it has been shown \cite{bradlyn_topological_2017,li_topological_2013} (see also Supplement \S\ref{sec:sup_topology}) that when the condition
\begin{equation}\label{topological criterion}
    t_{ss}+t_{pp} > \frac{|E_s-E_p|}{2}
\end{equation} 
is satisfied, the system is gapped and the lower band has a non-trivial Zak-Berry phase of $\pi$. It is also informative to examine the limit $t_{ss}=t_{sp}=t_{pp}=0$ and $E_s \neq E_p$, where the spectrum is gapped, and flat having band energies $E_{\pm}(k) = E_s$ or $E_p.$ This limit corresponds to momentum-independent eigenfunctions having vanishing Zak-Berry phase and Wannier functions localized on the atomic sites inside the unit cell. In the opposite limit where $t_{ss}=t_{sp}=t_{pp} = t \neq 0$ and $E_s=E_p$, the spectrum is also flat, with $E_{\pm}=\pm2t$. However, in this case the eigenfunctions have non-trivial Zak phases of $\pi$ and Wannier centers halfway between the unit cells implying a covalent insulator with $sp$-hybridized bonds. For a finite material having open boundaries, this placement of Wannier centers in a two-band insulator predicts the occurrence of protected fractional boundary charges at half-filling, and under suitable conditions, the existence of in-gap protected states \cite{hughes_inversion-symmetric_2011,Benalcazar_Quantization_2019,peterson_fractional_2020,li_fractional_2020}. These limits represent the extreme cases of the trivial and obstructed insulators, respectively, and a topological phase transition occurs whenever the condition in {Eq.~(\ref{topological criterion})} becomes an equality, i.e., when the system closes the energy gap, separating the two gapped phases on either side of the transition point.

\vspace{4pt}

\begin{figure}[p]
    \begin{adjustwidth*}{-1in}{-1in}
        \includegraphics[width=1.1\textwidth, clip=true, trim=0.3in 0in 0.2in 0in]{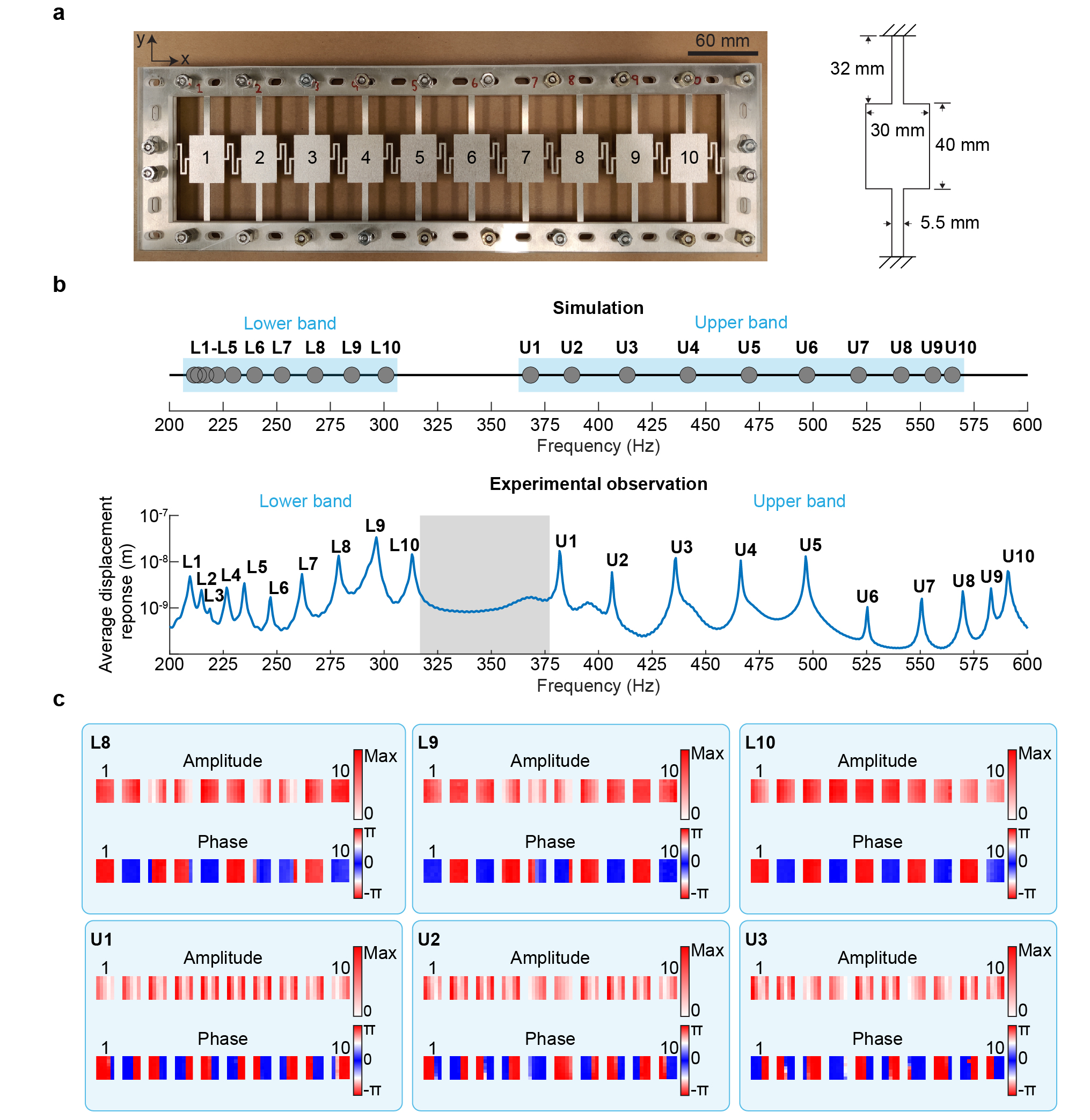}
    \centering
    \caption{
        \textbf{Experimental observations on the trivial structure.}
        \textbf{(a)} Photograph of the aluminum test structure, composed of ten resonators coupled via thin serpentine beams. The structure is clamped at the outer boundary using an acrylic frame.
        \textbf{(b)} The simulation predicted eigenfrequencies (above) and the SLDV-measured displacement response (below) averaged across the structure. The experimental spectrum shows a clear bandgap (shaded box) with 20 resolved modes distributed equally above and below the gap. 
        \textbf{(c)} Visualization of experimental data for a few modes measured near the bandgap. The amplitude colormap is normalized for each mode. In Supplement \S\ref{sec:sup_trivial}, we present visualizations for all 20 modes.
    }
    \label{trivial experiment}
    \end{adjustwidth*}
\end{figure}

\begin{figure}[p]
    \begin{adjustwidth*}{-1in}{-1in}

        \includegraphics[width=1.1\textwidth, clip=true, trim=0.3in 0in 0.3in 0in]{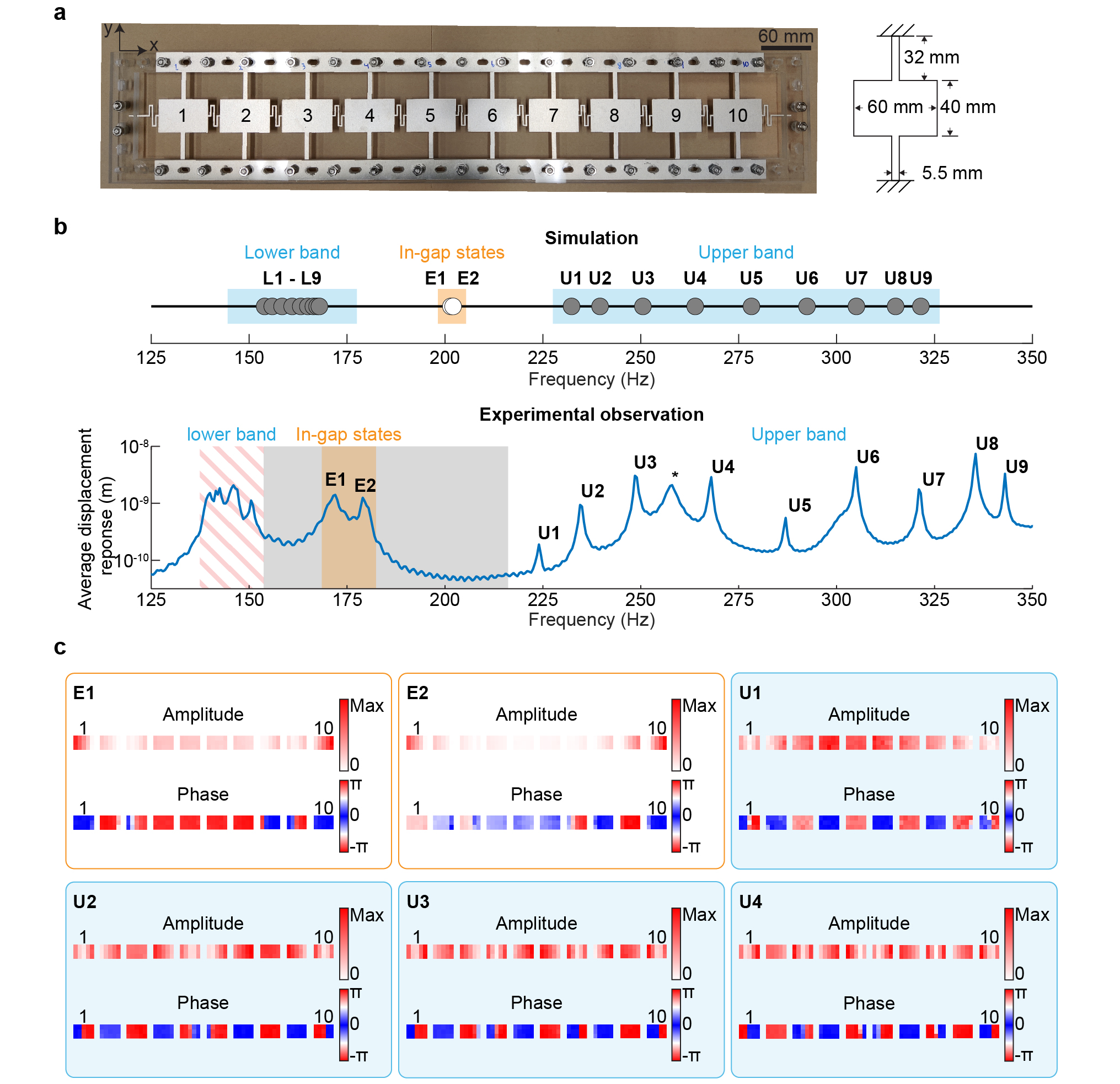} 
    \centering
    \caption{
        \textbf{Experimental observations on the non-trivial structure.}
        \textbf{(a)} Photograph of the aluminum test structure, composed of ten resonators in a manner similar to the trivial structure (Fig.~{\ref{trivial experiment}}).
        \textbf{(b)} The simulation predicted eigenfrequencies (above) and the SLDV-measured displacement response (below) averaged across the non-trivial structure. The experimental spectrum shows the expected bandgap, two bulk bands, and two in-gap states (labeled E1, E2 in orange box). The nine upper bulk modes can be identified individually while the lower bulk modes not individually resolvable due to its flatness (red stripes). The large amplitude response observed at 258 Hz (labeled *) is confirmed via SLDV measurement to be a spurious mode of the the frame. 
        \textbf{(c)} Visualization of experimental data for modes E1, E2, and a few modes near the bandgap. The in-gap modes are confirmed to be localized at the edges of the structure, as a confirmation of the topological protection and bulk-boundary correspondence. In Supplement \S\ref{sec:sup_topo}, we present visualizations for all the resolvable modes. 
        }        
    \label{topo experiment}
    \end{adjustwidth*}
\end{figure}

For our experimental implementation we employed mechanical plate resonators (Fig.~\ref{concept}c) whose two lowest  normal modes are a translational mode (even parity, $s$ orbital) and a torsional mode (odd parity, $p$ orbital). The effective location of the ``atom" is set by the location of the central tether attached to the plate. Because of this, it is not possible to cut or deform the structure to remove only one of these modes. Additionally, we chose this geometry as it pushes undesirable degrees of freedom to higher energies, allowing us to focus on only one $s$ and one $p$ orbital per atomic site. We performed finite element analysis (FEA) simulations using COMSOL Multiphysics to identify how manipulations of the dimensional parameters of the plate allow us to control the relative eigenfrequencies of the ODoF. Here, we chose the plate width $W$ as the tuning parameter and found that a modal energy crossing is achievable (Fig.~\ref{concept}d). With this foundation, we designed a 1D chain in which adjacent resonators are connected by long (serpentine wrapped) thin beams (Fig.~\ref{concept}e) that introduce non-zero hopping $t_{ss}$, $t_{pp}$, and $t_{sp}$ simultaneously. The hopping strengths are estimated with FEA simulations and we provide details on all of the dimensions of our structures in the Supplement \S\ref{sec:sup_details}. We can then identify the topologically non-trivial regime where Eq.~{\ref{topological criterion}} is satisfied. 

\vspace{4pt}

The bandstructure for our coupled-plate lattice can be calculated under periodic boundary conditions using FEA. The gap closing point representing the topological phase transition occurs for plate width $W=35$ mm (Fig.~{\ref{concept}}f). Reducing (increasing) the plate width separates the energies of the $s$ and $p$ orbital modes and opens a trivial (topologcal) bandgap as shown in Fig.~\ref{concept}g (Fig.~\ref{concept}h). Here we visualize the $s$ and $p$ components of the Bloch waves as a function of wave-vector in the first Brillouin zone by projecting the vibrational motion onto the even $s$-type and odd $p$-type orbital degrees of freedom (see Methods). For the trivial band gap the lower (upper) band is primarily $s$-type ($p$-type). Upon increasing $W$ to exceed the gap closing point (and hence satisfying Eq.~{\ref{topological criterion}}), the material acquires a non-trivial, inverted bandgap where the orbital character at $k=0$ switches as $k$ is swept across the Brillouin zone to $k=\pi/a$. The simulation under periodic boundary conditions shown in Fig.~{\ref{concept}}h confirms this through a clear band inversion at the boundary of the Brillouin zone. The change in the parity eigenvalues of the orbital states at the high symmetry points ($k=0$ and $k=\pi/a$) corresponds to Zak phase of $\pi$ \cite{hughes_inversion-symmetric_2011,turner_quantized_2012}.

\vspace{4pt}

Our experiments were performed using structures made by waterjet cutting single-pieces of thin aluminum sheets (details in Supplement \S\ref{sec:sup_details}). We used arrays of ten resonators/plates to test both the trivial (Fig.~{\ref{trivial experiment}}a) and non-trivial cases (Fig.~{\ref{topo experiment}}a). To identify the spectrum and modes the material was excited with a loudspeaker using periodic chirp signals while the out-of-plane displacement response across the surface of the array was measured with a scanning laser Doppler vibrometer (Polytec PSV-500, SLDV). The data thus obtained contains spectral information on the local vibration amplitude and phase relative to the excitation signal, i.e., we extract a complex value $\tilde{a}(x, y)$ across the structure where $x$ and $y$ are the spatial coordinates labeling points on the array surface (Fig.~{\ref{trivial experiment}}a and Fig.~{\ref{topo experiment}}a). Additional details on the experimental setup are available in the Supplementary Material \S\ref{sec:sup_details}.   

\vspace{4pt}

Experimental measurements of the averaged displacement response integrated over the trivial structure ($\frac{1}{\textrm{area}} \iint_\textrm{all resonators} |\tilde{a}(x, y)|$) are presented in Fig.~{\ref{trivial experiment}}b, and we find that the observed mode frequencies are closely matched to FEA simulations. In Fig.~{\ref{trivial experiment}}c we present the SLDV-measured mode shapes -- with $|\tilde{a}(x, y)|$ and phase $\angle \tilde{a}(x, y)$ plotted separately -- for a few selected modes near the band edge. A full display of all twenty measured modes is provided in the Supplement \S\ref{sec:sup_trivial}. A brief visual inspection of the modes near the band edge suggests that the upper band exhibits $p$-type characteristics while the lower band exhibits $s$-type characteristics, though we formally analyze this below.

\vspace{4pt}

We proceed by applying the same experimental procedure to the non-trivial structure, with measurements presented in Fig.~{\ref{topo experiment}}b. Once again, we are able to identify the upper and lower bulk bands and confirm that the spectral frequencies and the mode shapes of the measured modes matches that from the simulations. The upper band modes can be individually identified, but we cannot individually resolve the modes in the lower band because they are more densely packed since the lower band has a much flatter dispersion (Fig.~{\ref{concept}}h). Also, distinct from the trivial case, a pair of in-gap states are observed. In Fig.~{\ref{topo experiment}}c, we display the in-gap modes and the first few upper band modes (a complete display of all 11 resolvable modes is presented in the Supplement \S\ref{sec:sup_topo}). Our data shows that the two in-gap modes E1 and E2 are spatially localized at the array edges. A rough visual inspection of the upper band modes indicates a mixture of $s$ and $p$ characteristics. Most notably, the mode that lowest energy mode of the upper band has strong $s$-type character, as was predicted in Fig.~{\ref{concept}}h--indicating a band inversion. We will formally confirm this below.

\begin{figure}[t]
  \begin{adjustwidth*}{-1in}{-1in}
    \hsize=\linewidth
        \includegraphics[width=1.4\textwidth, clip=true, trim=0.5in 0in 0.5in 0in]{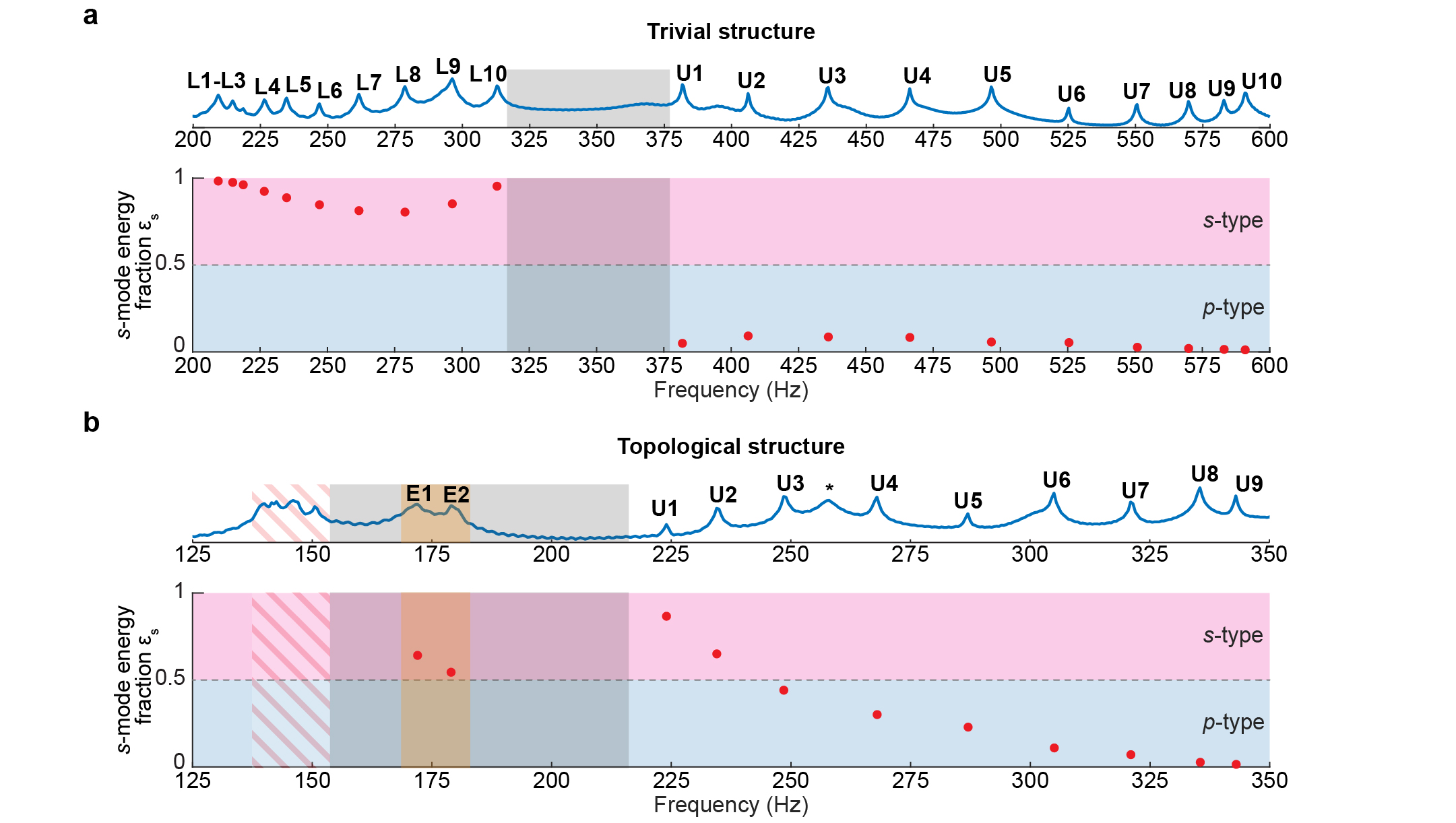}
    
    \centering
    \caption{
       \textbf{Calculation of the $s$-mode energy fraction $\epsilon_s$ for all the experimentally observed resolvable modes in Fig.~{\ref{trivial experiment} and Fig.~{\ref{topo experiment}}}.}
        \textbf{(a)} $\epsilon_s$ calculated for the topologically trivial array shows lower band modes with strong $s$-type character separated from the $p$-type upper band modes.
       \textbf{(b)} $\epsilon_s$ calculated for the topologically non-trivial array shows a gradual transition from $s$-type bulk modes to $p$-type ones in the upper band spectrum. Additionally, the in-gap edge modes are $sp$-hybridized. Both observations agree with the prediction offered by the simulation in Fig.~\ref{concept}g,h.
    }
    \label{s-energy fraction}
   \end{adjustwidth*}
\end{figure}

\vspace{4pt}

To formally quantify the band inversion in the topological structure, we project the experimentally measured displacement functions onto the $s$ and $p$ ODoFs. For the $n^\textrm{th}$ resonator we project the measured displacement (see Methods) onto each ODoF to obtain the energy distribution $\mathcal{E}_{s, n}$ and $\mathcal{E}_{p, n}$. The total contributions from each ODoF are then defined as $\mathcal{E}_{s, \textrm{tot}} = \sum_{\forall n} \mathcal{E}_{s, n}$ and $\mathcal{E}_{p, \textrm{tot}} = \sum_{\forall n} \mathcal{E}_{p, n}$. Finally, we use the $s$-type ODoF energy fraction as a normalized metric for the contributions by defining $\epsilon_s = \mathcal{E}_{s, \textrm{tot}} / (\mathcal{E}_{s, \textrm{tot}} + \mathcal{E}_{p, \textrm{tot}})$. Here we have made an implicit assumption that all the mechanical energy is stored in either the $s$-type or $p$-type modes since other modes are far from the frequency range of interest. For $\epsilon_s > 1/2$ the selected mode has a dominant $s$-type character, while $\epsilon_s < 1/2$ indicates a $p$-type character. We now apply the $\epsilon_s$ metric on each of the resolvable modes that are measured in our experiments (Fig.~{\ref{s-energy fraction}}). For the trivial structure (Fig.~{\ref{s-energy fraction}}a), we find that all bulk modes in the lower band exhibit a clear $s$-type character, whereas the modes in the upper band are all $p$-type, matching the prediction in (Fig.~{\ref{concept}}g). On the other hand, the upper band of the non-trivial structure (Fig.~{\ref{s-energy fraction}}b) exhibits a gradual shift from $s$-type to $p$-type modal character directly confirming the band inversion and the formation of $sp$-hybrid orbital modes predicted in Fig.~{\ref{concept}}h. 

\vspace{4pt}

The ability to create artificial atoms in synthetic materials can unlock new venues for investigating orbital physics--allowing emulation of fermionic lattices as well as novel bosonic quantum matter \cite{bloch_many-body_2008,li_physics_2016}. Recently, it has been emphasized theoretically\cite{bradlyn_topological_2017} and demonstrated experimentally \cite{schulz_photonic_2022,zhang_realization_2023} the key role that orbital symmetry representations play in creating topologically non-trivial bandstructures. Our work builds upon this: by collapsing all the SDoF within a unit cell to a single point to create an artificial mono-atomic lattice with multiple ODoF at each atomic site, we have realized a one dimensional TCI with robust symmetry representations of ODoF in each unit cell.  A key feature of our mechanical experimental realization is the artificial atoms capable of hosting multiple orbital modes while simultaneously enabling easy control of the relative energies between these modes. This ability to involve higher orbital degrees of freedom offers the opportunity to study novel topological phenomena \cite{li_physics_2016}. With good scalability and control over the relative modal energies, the mechanical orbital-based platform can be used to build systems with nearly flatbands with non-trivial topology to study controlled transport and localization in lattice systems \cite{sun_nearly_2011,jacqmin_direct_2014,milicevic_type-iii_2019,caceres-aravena_perfect_2019,caceres-aravena_topological_2020,caceres-aravena_Controlled_2022} or topological semimetals in two dimensions by mixing of $p$ and $d$ orbitals \cite{li_physics_2016,sun_2012_topological}. Additionally, we expect to see realizations of unconventional higher-order topological phases unique to systems allowing orbital hybridization \cite{Mazanov_tailoring_2022,Gladstein-Gladstone_spin_2022}.

\section*{Methods}

\textbf{Projection onto the orbital basis -- } We define an energy-normalized orbital basis $s_\textrm{norm}(x,y)$ and $p_\textrm{norm}(x,y)$, i.e. the displacement eigenfunctions for one resonator, by setting $\iint_{x,y} | s_\textrm{norm}(x,y) |^2 = \iint_{x,y} | p_\textrm{norm}(x,y) |^2 = 1$. We then use a projection operation to evaluate how the energy in any displacement profile having complex amplitude $\tilde{a}(x,y)$ is distributed between the $s$ and $p$ ODoFs for that resonator, by evaluating the energies $\mathcal{E}_s = |\iint_{x,y} \tilde{a}(x,y) \cdot s_\textrm{norm}(x,y) |^2$ and $\mathcal{E}_p = |\iint_{x,y} \tilde{a}(x,y) \cdot p_\textrm{norm}(x,y) |^2$.

\section*{Author contributions}

J.N., W.A.B., G.B., G.L., and V.W. conceived the system and its mechanical metamaterial implementation. V.W. and G.L. conducted the experiments. G.L., J.N., V.W., G.B. and T.L.H. analysed the experimental results. Y.M. and T.L.H. provided additional theory support. G.B. supervised the effort. All authors contributed to the writing of the manuscript.

\section*{Acknowledgments}

This work was sponsored by the Multidisciplinary University Research Initiative (MURI) grant N00014-20-1-2325.
The authors express their gratitude to Osama Jameel from Polytec Inc. for guidance and assistance on the scanning vibrometry measurements in this work.
J. N. acknowledge support from the US Department of Energy, Office of Basic Energy Sciences, Division of Materials Sciences and Engineering. This work was performed, in part, at the Center for Integrated Nanotechnologies, an Office of Science User Facility operated for the U.S. Department of Energy (DOE) Office of Science. Sandia National Laboratories is a multimission laboratory managed and operated by National Technology $\&$ Engineering Solutions of Sandia, LLC, a wholly owned subsidiary of Honeywell International, Inc., for the U.S. DOE’s National Nuclear Security Administration under contract DE-NA-0003525. This paper describes objective technical results and analysis. Any subjective views or opinions that might be expressed in the paper do not necessarily represent the views of the U.S. Department of Energy or the United States Government.

\FloatBarrier

\newpage

\renewcommand*{\citenumfont}[1]{S#1}
\renewcommand*{\bibnumfmt}[1]{[S#1]}
\newcommand{\beginsupplement}{%
        \setcounter{table}{0}
        \renewcommand{\thetable}{S\arabic{table}}%
        \setcounter{figure}{0}
        \renewcommand{\thefigure}{S\arabic{figure}}%
        \setcounter{equation}{0}
        \renewcommand{\theequation}{S\arabic{equation}}
        \setcounter{section}{0}
        \renewcommand{\thesection}{S\arabic{section}}%
}

\beginsupplement

\begin{center}

{\Large \textbf{Supplementary Information: \\ A mono-atomic orbital-based\\ 1D topological crystalline insulator}}

\vspace{12pt}
{ Gengming Liu $^{1\dag}$, Violet Workman $^{1\dag}$, Jiho Noh$^{2,3\dag}$, Yuhao Ma$^{1}$, \mbox{Taylor L. Hughes$^{1}$,} Wladimir A. Benalcazar$^4$, Gaurav Bahl$^{2*}$ \\}
\vspace{12pt}
\small{$^1$ Department of Physics, $^2$ Department of Mechanical Science $\&$ Engineering, \\
    University of Illinois at Urbana–Champaign, Urbana, IL 61801 USA \\
    $^3$ Sandia National Laboratories, Albuquerque, NM 87185, USA. \\
    $^4$ Department of Physics, Emory University, Atlanta, GA 30322, USA \\}
\end{center}

\vspace{12pt}

\section{Topology of the orbital-based mono-atomic 1D TCI model}\label{sec:sup_topology}
\setcounter{page}{1}

This section of our supplement is dedicated to exploring the topological aspects of the orbital model depicted in Fig.~\ref{concept}b in the main text. We commence with the derivation of the Bloch Hamiltonian from the real space tight-binding model of the system.
 
\begin{equation}\label{eqs1}
H=\sum_{n}\Psi^\dag(n)[U\Psi(n)+V\Psi(n+1)+V^\dag\Psi(n-1)], 
\end{equation}

\begin{equation}
    \label{eqs2}
        U=
        \begin{bmatrix}
            E_s&&0\\
            0&&E_p\\
        \end{bmatrix},\quad 
        V=\begin{bmatrix}
            t_{ss}&&t_{sp}\\
            -t_{sp}&&-t_{pp}\\
        \end{bmatrix},
\end{equation}
where n is the unit cell index, $E_{s/p}$ are the orbital energies and $\Psi(n)$ the spinor
\begin{equation*}
        \Psi(n)=
        \begin{bmatrix}
            \C_{s}(n)\\
            \C_{p}(n)\\
        \end{bmatrix}.
\end{equation*}
$\C_{s}(n)$ and $\C_{p}(n)$ are the annihilation operator of $s$ and $p$ orbital mode on atom site $n$ respectively. Applying a Fourier transform to the operator $\C_{s/p}(n) = \frac{1}{\sqrt{N}}\sum_{k} e^{ikn} \C_{s/p}(k) $, we have

\begin{equation}\label{eqs3}
H=\sum_{k} \Psi^\dag(k)H(k)\Psi(k),
\end{equation}
where $k$ is the wave vector and the Bloch Hamiltonian in the orbital basis is
\begin{equation}\label{eqs4}
H(k)=U+Ve^{ika}+V^\dag e^{-ika}=
        \begin{bmatrix}
            E_s+2t_{ss}\cos{ka}&&2it_{sp}\sin{ka}\\
            -2it_{sp}\sin{ka}&&E_p-2t_{pp}\cos{ka}\\
        \end{bmatrix},
\end{equation}
just as presented in Eq.~\ref{Hamiltonian} of the main text. We can visualize the trajectory traced out by the wavefunction across the Brillouin zone by expanding Eq.~\ref{eqs4} in terms of Pauli matrices defined as
\begin{equation*}\hat\sigma_x=\begin{bmatrix}
0 & 1 \\
1 & 0 
\end{bmatrix},\hspace{2mm} \hat\sigma_y=\begin{bmatrix}
0 & -i \\
i & 0 
\end{bmatrix},\hspace{2mm}  \hat\sigma_z=\begin{bmatrix}
1 & 0 \\
0 & -1 
\end{bmatrix}.
\end{equation*}
The result is Eq.~\ref{Hamiltonian in Pauli matrices} from the main text in the form of $H(k)=h_0\mathbb{I}+\vec{h}(k)\cdot\vec{\sigma}$ and the topological winding of $\vec{h}(k)$ across the Brillouin zone gives the Zak phase of the system \cite{li_topological_2013S}.

The topology of our orbital model can be readily shown by analyzing the following two limits \cite{bradlyn_topological_2017S}. The first one is when $t_{ss}=t_{sp}=t_{pp}=0$ and $E_s\neq E_p$, the system has two flat bands of energy $E_s$ and $E_p$ with corresponding eigenstates:

\begin{equation*}
\ket{E_s} =\begin{pmatrix}
1 \\
0 
\end{pmatrix},\hspace{4mm} \ket{E_{p}}=\begin{pmatrix}
0 \\
1 
\end{pmatrix},
\end{equation*} 
both having Zak phases of 0. Therefore, the Wannier functions are localized at the center of the unit cell--directly on the atom. The occupied Wannier functions are thus pure $s$-type (or $p$-type, depending on their relative modal energies) orbital modes sitting on the atomic sites.

The other limit is reached when $t_{ss}=t_{sp}=t_{pp}=t$ and $E_s= E_p = E $. The system, again, has two flat bands
\begin{equation*}
E_{\pm}= E \pm 2t.
\end{equation*}
However the valence and conduction bands
\begin{equation*}
\psi_-(k)=e^{ika/2}\begin{pmatrix}
1+\cos{ka} \\
i\sin{ka}
\end{pmatrix},\hspace{4mm} \psi_+(k)=e^{ika/2}\begin{pmatrix}
1-\cos{ka} \\
i\sin{ka}
\end{pmatrix},
\end{equation*} 
are now no longer trivial with quantized Zak phase of
\begin{equation}\label{eqs6}
\phi_{\pm}=i\int_{-\pi}^{\pi}dk\,\psi_\pm^\dag\, \nabla_k\,\psi_\pm=\pi.
\end{equation}
From this, we recognize that the valance and conduction band Wannier functions must be localized half-way between atomic sites.
This localization of Wannier functions is distinct from that of the trivial atomic limit and represents a covalent obstructed atomic insulator. A topological phase transition between these two limiting cases occurs when the condition in Eq.~\ref{topological criterion} becomes an equality, and hence the bandgap closes. Further reduction of the orbital energy difference, i.e., whenever Eq.~\ref{topological criterion} is satisfied, encourages the formation of $sp$-hybridized orbitals to create a band inversion. The resulting Wannier functions sit between the atomic sites and generate a non-trivial bulk charge polarization. At the boundary between materials with the distinct Zak phase, we expect bound charge and boundary states to appear \cite{shockley_surface_1939S}.

We can also compare ODoF model with the related SDoF system represented by the Su-Schrieffer-Heeger (SSH) model \cite{su_solitons_1979S}:

\begin{equation}\label{eqs7}
    \begin{aligned}
        H_{SSH}(k)=&
        \begin{bmatrix}
            0&&\nu+e^{ika}\omega\\
            \nu+e^{-ika}\omega&&0\\
        \end{bmatrix},
        \\= & -(\omega\sin{ka})\hat\sigma_y+    (\nu+\omega\cos{ka})\hat\sigma_x,
    \end{aligned}
\end{equation}
where $H_{ SSH}(k)$ is the Bloch Hamiltonian written in sublattice basis, $\nu$ is the intra-cell coupling strength, and $\omega$ the inter-cell strength. We can represent the odd and even parities of the orbital bases with the phases of the two sublattices $A$ and $B$:
\begin{equation}\label{eqs8}
\begin{bmatrix}
S\\P
\end{bmatrix}=
\frac{\sqrt{2}}{2}
\begin{bmatrix}
1&&1\\
-1&&1
\end{bmatrix}
\begin{bmatrix}
A\\B    
\end{bmatrix}.
\end{equation}
We then observe that the unitary matrix $G$ in Eq.~\ref{eqs8} is equivalent to a rotation of $-\frac{\pi}{4}$. A change of basis can then be performed on Eq.~\ref{eqs7} to bring the SSH Bloch Hamiltonian into orbital basis:
\begin{equation}\label{eqs9}
\begin{aligned}
G^{-1}\mathcal{H}G=& \frac{1}{2}
\begin{bmatrix}
1&&-1\\
1&&1\\
\end{bmatrix}
\begin{bmatrix}
0&&\nu+e^{ika}\omega\\
\nu+e^{-ika}\omega&&0\\
\end{bmatrix}
\begin{bmatrix}
1&&1\\
-1&&1\\
\end{bmatrix},
\\= & 
\begin{bmatrix}
-\nu-\omega\cos{ka}&&i\omega\sin{ka}\\
-i\omega\sin{ka}&&\nu+\omega\cos{ka}\\
\end{bmatrix},
\\= & -(\omega\sin{ka})\hat\sigma_y - (\nu+\omega\cos{ka})\hat\sigma_z.
\end{aligned}
\end{equation}
Comparing Eq.~\ref{eqs9} with the last two terms of Eq.~\ref{Hamiltonian in Pauli matrices}, we see that the SSH model is mapped to the orbital model with a change of basis \cite{suppression_2022S,caceres-aravena_perfect_2019S,caceres-aravena_topological_2020S,li_topological_2013S}.
\FloatBarrier   %

\vspace{24pt}

\section{Robustness to disorder}
\label{sec:sup_robustness_to_onsite}

Here we explore the robustness of the orbital-based TCI model with a direct comparison to the well-known SSH chain. We leverage the \textit{fractional sector charge} as a quantitative metric to compare the two cases. 

\begin{figure}[b!]
    \begin{adjustwidth*}{-1in}{-1in}
    \hsize=\linewidth
    \includegraphics[width=1.2\textwidth]{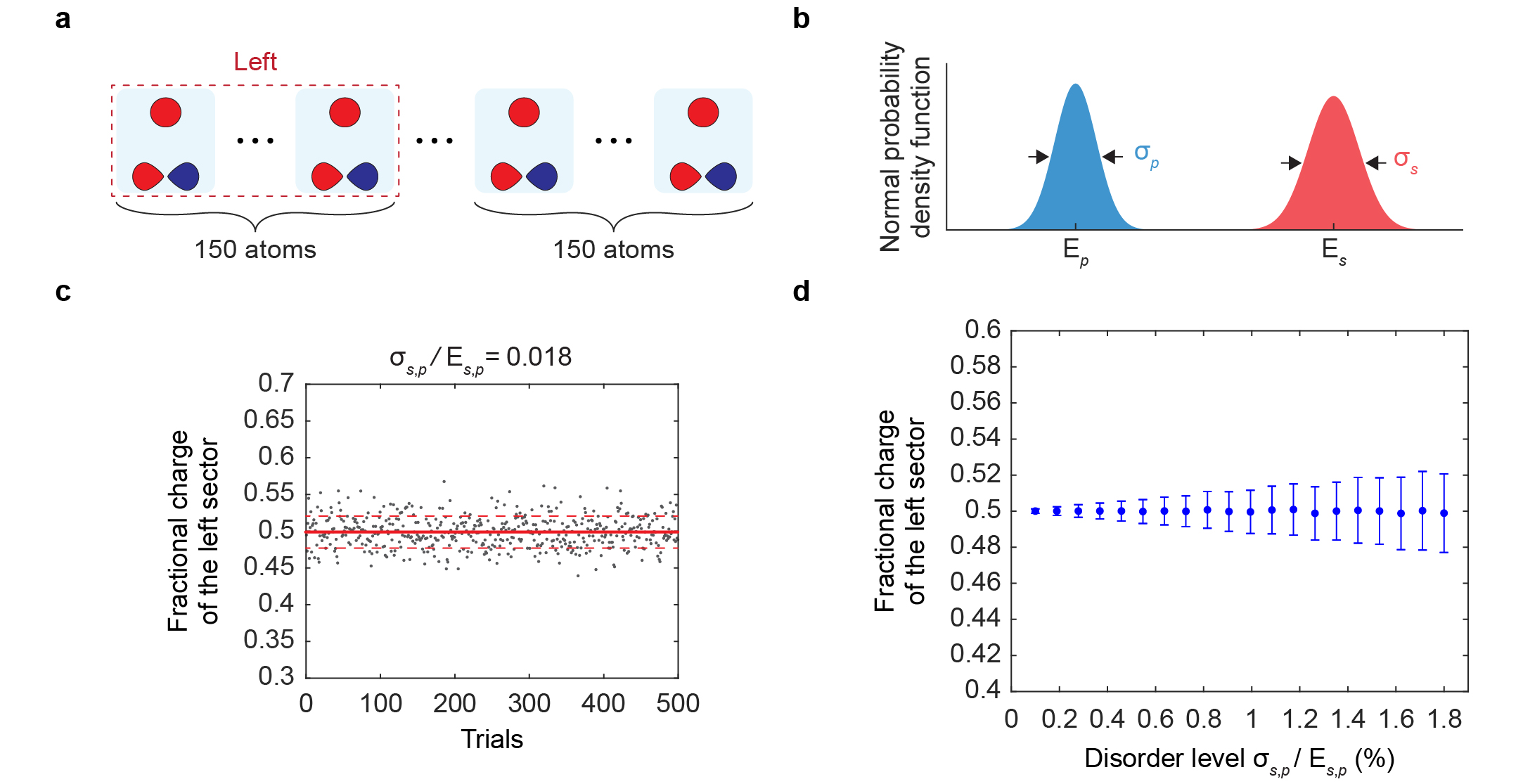}
    \centering
        \caption{
        \textbf{Calculation of the fractional sector charge for the left half of an orbital-based mono-atomic 1D TCI with disordered orbital energies.}
        \textbf{(a)} We are interested in the fractional sector charge in the left sector for a system containing 300 atoms.
        \textbf{(b)} The orbital energies at each atom are sampled at random from a normal distribution whose standard deviation $\sigma_s \propto E_s$ ($\sigma_p \propto E_p$) characterizes the disorder level.
        \textbf{(c)} An example calculation of the fractional sector charge calculated over 500 trials at the disorder level $\sigma_{s,p}/E_{s,p}=0.018$. The average value (thick red line) and the standard deviation from the average (dashed red lines) are identified.  
        \textbf{(d)} Statistics from fractional sector charge calculations performed at increasing disorder levels. Data points denote the mean values while the error bars show the standard deviation.
        }
    \label{fig:S_robust_orbital}
    \end{adjustwidth*}
\end{figure}

We begin by considering the general case of $C_{n}$-symmetric insulators, i.e. systems that can be subdivided into $n$ rotationally-symmetric sectors. The integral of the spatial charge density (under some filling choice, e.g. half filling) performed over an entire sector is defined as the \textit{sector charge} \cite{peterson_fractional_2020S}. Due to symmetry, the sector charge is identical for each sector, but when accumulated over all sectors the total charge must always take an integer value $\mathcal{N}$. Thus, the fractional part of the charge per sector (the \textit{fractional sector charge}) is defined as $\mathcal{N}/n \ \operatorname{mod}\ 1$ and must be quantized in units of $1/n$. Since, in this exploration, we are working with 1-dimensional (1D) materials constrained by $C_2$ or mirror symmetry, the fractional sector charge must be quantized to units of $1/2$. Importantly, the fractional sector charge cannot be changed by symmetric adiabatic deformations and conveys the filling anomaly for TCIs \cite{Benalcazar_Quantization_2019S,peterson_fractional_2020S}. Moreover, the fractional sector charge can help classify some crystalline topological phases -- i.e. if a non-trivial phase has a fractional sector charge of $1/2$ (or $0$) then the trivial phase will have fractional sector charge of $0$ (or $1/2$). Deviations from a quantized sector charge due to disorder can be used as a quantitative metric of the robustness of the topological phase. 

\begin{figure}[t!]
    \begin{adjustwidth*}{-1in}{-1in}
    \hsize=\linewidth
    \includegraphics[width=1.2\textwidth]{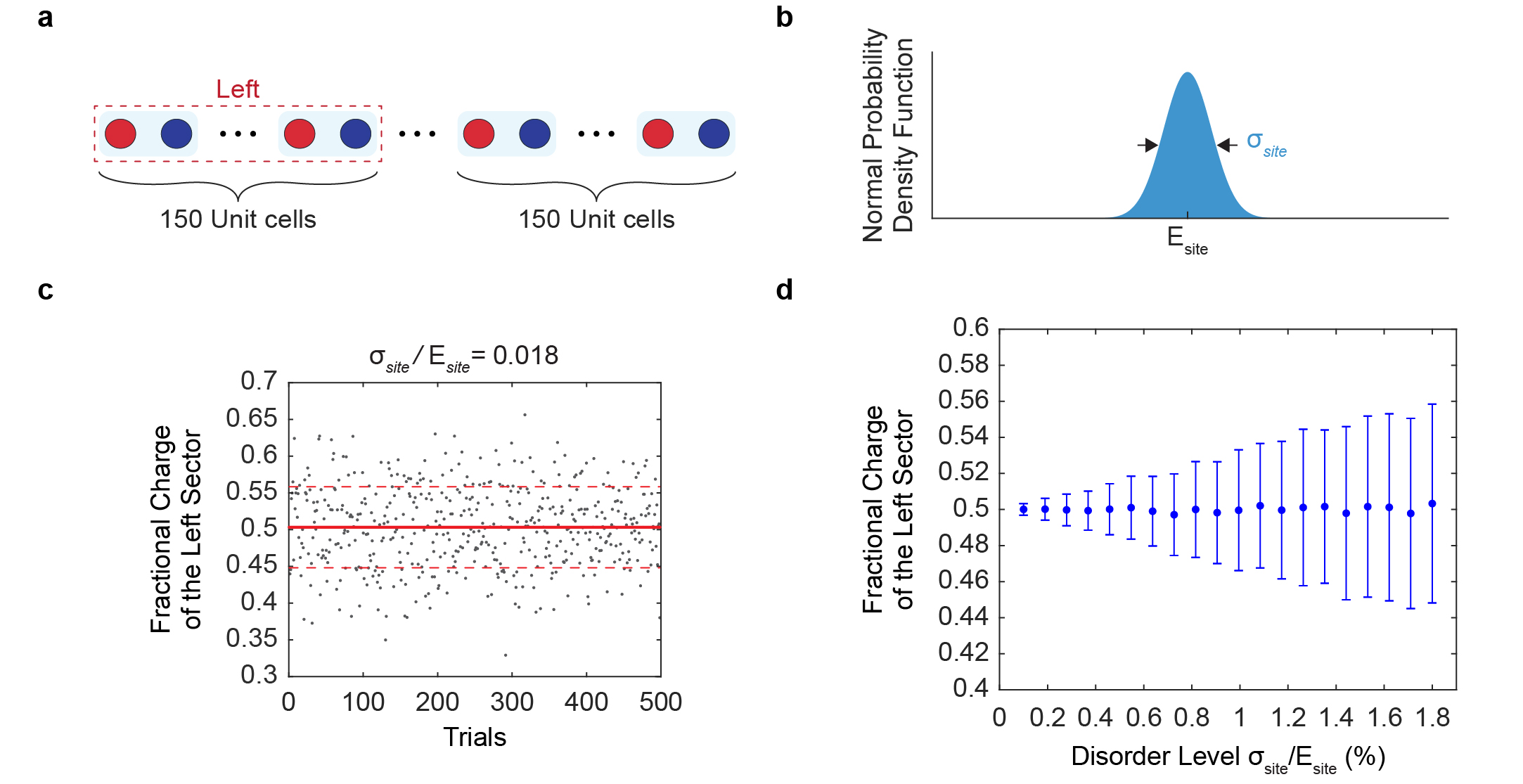}
    \centering
        \caption{
        \textbf{Calculation of the fractional sector charge for the left half of a SSH 1D TCI with disordered on-site energies.}
        \textbf{(a)} Similar to Fig.~\ref{fig:S_robust_orbital} we are interested in the fractional sector charge in the left sector for a system containing 300 unit cells.
        \textbf{(b)} The on-site energies are sampled at random from a normal distribution whose standard deviation $\sigma_{site}\propto E_{site}$ characterizes the disorder level.
        \textbf{(c)} An example calculation of the fractional sector charge calculated over 500 trials at the disorder level $\sigma_{site}/E_{site}=0.018$.
        \textbf{(d)} Statistics from fractional sector charge calculations performed at increasing disorder levels. Results are presented at the same scale as Fig.~\ref{fig:S_robust_orbital}d and convey the much higher sensitivity to disorder than the orbital-based TCI.
        }
    \label{fig:S_robust_ssh}
    \end{adjustwidth*}
\end{figure}

As described in the main text, we are interested in testing the robustness of the orbital-based mono-atomic 1D TCI ({Fig.~\ref{fig:S_robust_orbital}a}) to disorder of the orbital energies. This test is performed using a tight binding calculation of a 1D TCI containing 300 atoms (600 total orbital degrees of freedom) to which we introduce normally-distributed random disorder to $E_s$ and $E_p$ ({Fig.~\ref{fig:S_robust_orbital}b}). The standard deviations of the spread $\sigma_s$ and $\sigma_p$ are scaled to the respective orbital energies so that the fractional disorder is the same. The fractional sector charge for the left sector ({Fig.~\ref{fig:S_robust_orbital}a}) is then evaluated using 500 trials at each disorder level quantified by $\sigma_{s,p} / E_{s,p}$. One example set of 500 simulations for $\sigma_{s,p}/E_{s,p} = 0.018$ is shown in {Fig.~\ref{fig:S_robust_orbital}c}. The overall mean and standard deviation of fractional sector charge is then evaluated as a function of the level of disorder, and presented in Fig.~\ref{fig:S_robust_orbital}d. As expected, we see a gradual increase in the deviation of fractional sector charge as the disorder level is increased.
     
We next perform a similar analysis on a non-trivial SSH tight-binding model ({Fig.~\ref{fig:S_robust_ssh}a}) containing 300 unit cells (600 total spatial degrees of freedom). We again introduce normally-distributed random disorder to the on-site energies ({Fig.~\ref{fig:S_robust_ssh}b}) and perform 500 simulations at each disorder level. The addition of on-site disorder in this case breaks the chiral symmetry, and the left fractional sector charge  ({Fig.~\ref{fig:S_robust_ssh}a}) of the model should change continuously with the level of disorder. An example set of simulations with disorder level matched to {Fig.~\ref{fig:S_robust_orbital}c} is presented in {Fig.~\ref{fig:S_robust_ssh}c}, from which we can immediately see the greater spread in the charge quantization. The overall statistics of the fractional sector charge are presented in {Fig.~\ref{fig:S_robust_ssh}d}. When compared against {Fig.~\ref{fig:S_robust_orbital}d}, these results convey the higher robustness of the orbital-based 1D TCI to perturbation of energies.

\FloatBarrier

\newpage  

\section{Details on the plate mechanical material implementation}
\label{sec:sup_details}

\begin{figure}[b!]
    \begin{adjustwidth*}{-1in}{-1in}
    \hsize=\linewidth
    \includegraphics[width=0.9
    \textwidth]{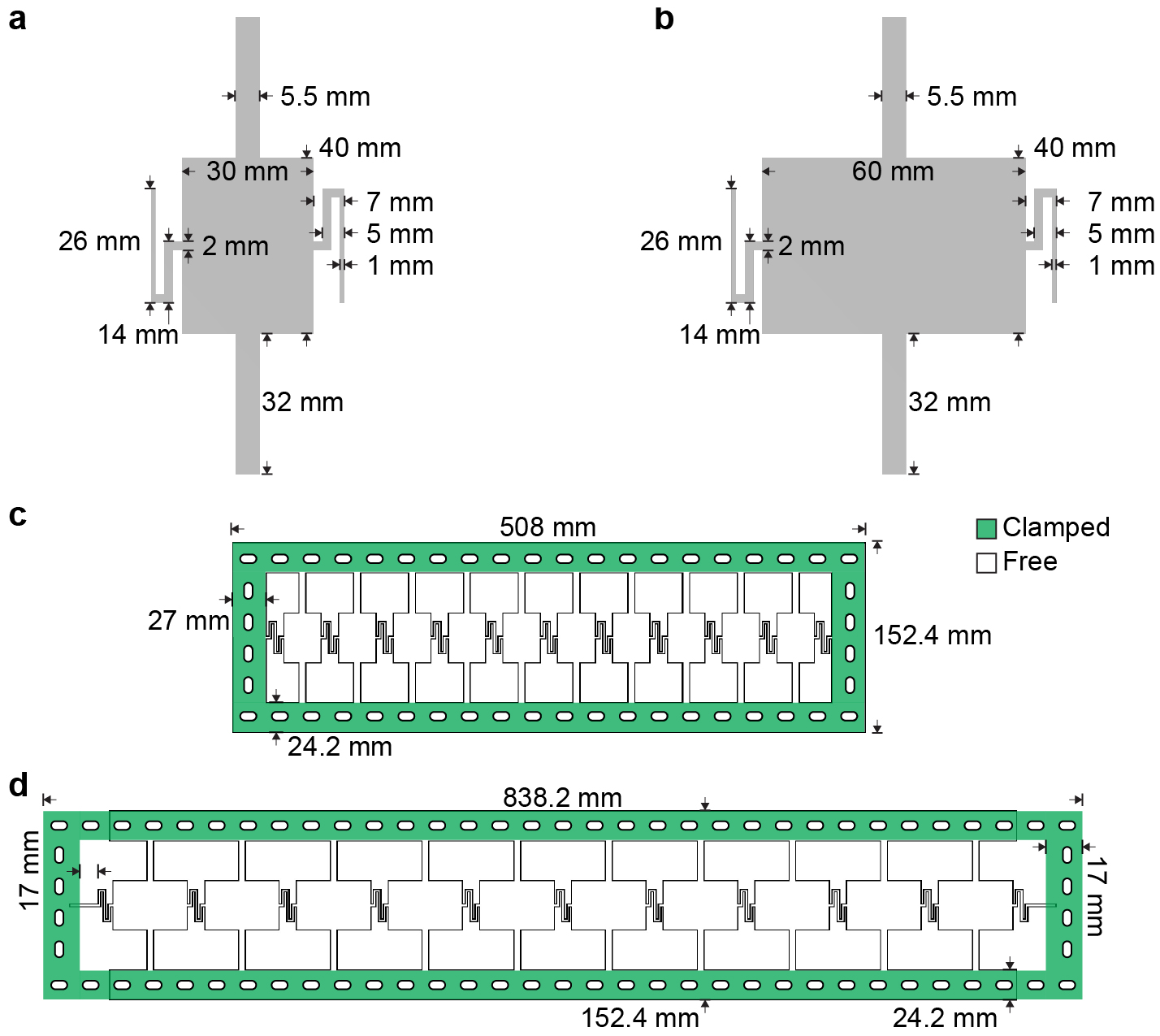}
    \centering
    \caption{
        \textbf{All relevant dimensions for the trivial and the non-trivial structures.}
        \textbf{(a)} Resonator and tether dimensions for a trivial unit cell.
        \textbf{(b)} Resonator and tether dimensions for a non-trivial unit cell.        %
        \textbf{(c)} Trivial chain dimensions. Green-shaded regions are sandwiched by acrylic sheets for anchoring.
        \textbf{(d)} The same for the non-trivial chain.}
    \label{fig:s_dimensions}
    \end{adjustwidth*}
\end{figure}

\begin{figure}[t]
    \begin{adjustwidth*}{-1in}{-1in}
    \hsize=\linewidth
    \includegraphics[width=0.9\textwidth]{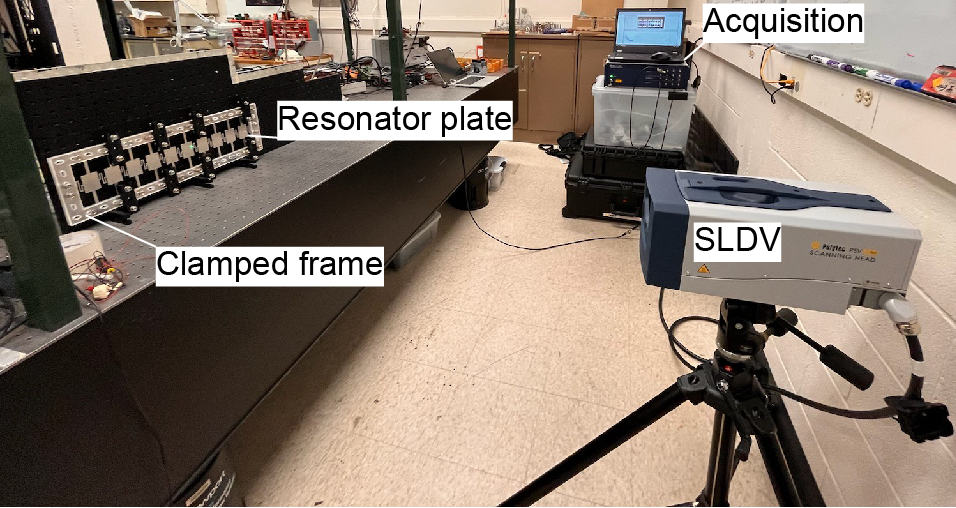}
    \centering
    \caption{
        \textbf{Experimental setup with the SLDV.}
    }
    \label{fig:s_exp_steup}
    \end{adjustwidth*}
\end{figure}

Both of the mechanical resonator arrays used in experiment are waterjet cut into their final shapes in one-shot on 6061 aluminum plates (0.79 mm thickness). Here, we show all dimensions of the trivial (Fig.~\ref{fig:s_dimensions}a) and the non-trivial (Fig.~\ref{fig:s_dimensions}b) unit cell. Overall dimensions of the trivial (Fig.~\ref{fig:s_dimensions}c) and non-trivial (Fig.~\ref{fig:s_dimensions}d) materials are also shown. Two acrylic sheets (6.35 mm thickness) are also cut to partially (shaded green) sandwich the aluminum layer--providing anchor points for the resonator spring tethers while leaving room for the paddle plates to vibrate freely. 
The three layers are then secured with screws and fixed onto an optical breadboard set vertically on an optical table--allowing access for vibrometry measurements from the side of the table [Fig.~\ref{fig:s_exp_steup}].

The non-trivial structure has longer coupling tethers at array ends to offset the loading effects such that the topological edge states appear near the middle of the bandgap. A SLDV (Polytec PSV-500) is then used to measure the out-of plane displacement response of the structures while a loudspeaker is used to excite the structure as described in the main text. The sensitivity of the SLDV is set to 12.5 micrometers per volt signal output and 35(15) sample points across each paddle pate from the trivial(non-trivial) material are measured in experiment.

\FloatBarrier   %

\newpage
\section{Visualization of all bulk modes for the trivial structure}
\label{sec:sup_trivial}
We show all the bulk modes of the trivial structure with side-to-side comparison to the eigenmode shapes predicted by the FEA software COMSOL. We see that the measured mode shapes matches very well with the simulation and this agreement also helps with the identification of the bandgap.

\begin{figure}[htp]
    \begin{adjustwidth*}{-1in}{-1in}
    \hsize=\linewidth
    \includegraphics[width=1.3\textwidth]{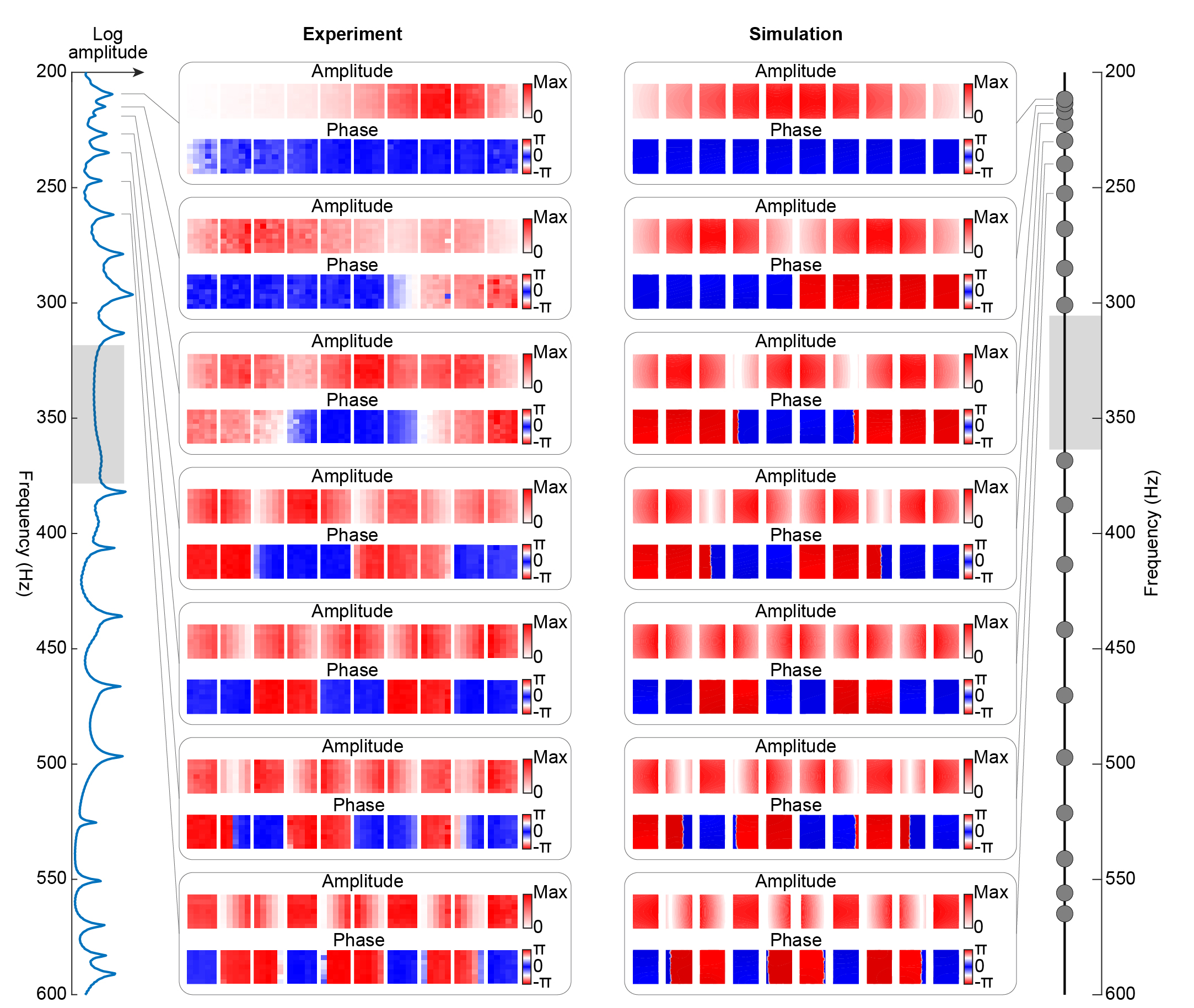}
    \centering
    \caption{
        \textbf{Bulk modes (L1 to L7) for the trivial structure.}
        We show the measured (left half) and the simulated (right half) mode shapes for the bulk mode L1 through L7.
    }
    \label{fig:s_trivial_bulk_modes_1/3}
    \end{adjustwidth*}
\end{figure}

\begin{figure}[htp]
    \begin{adjustwidth*}{-1in}{-1in}
    \hsize=\linewidth
    \includegraphics[width=1.3\textwidth]{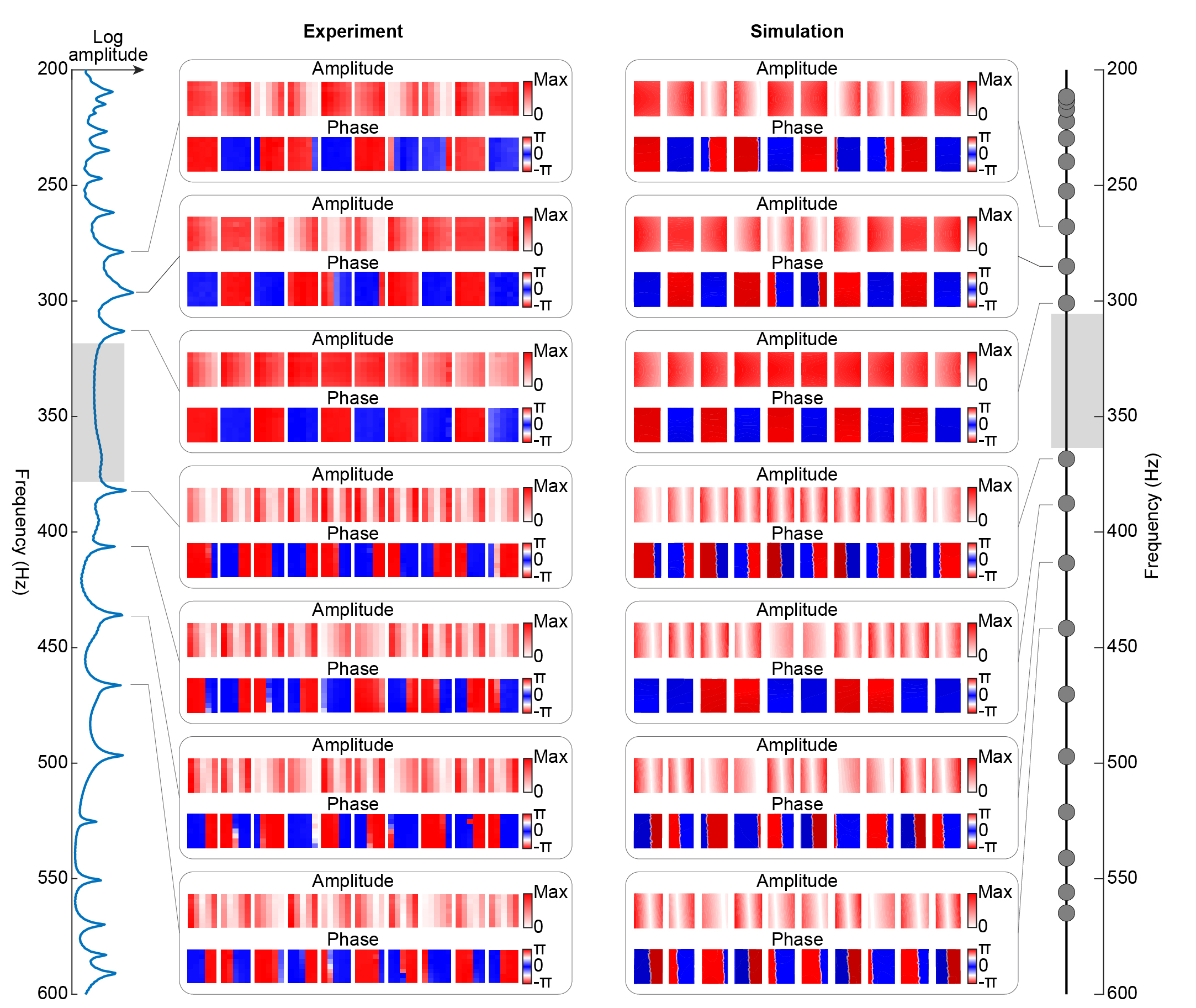}
    \centering
    \caption{
        \textbf{Bulk modes (L8 to U4) for the trivial structure.}
        We show the measured (left half) and the simulated (right half) mode shapes for the bulk mode L8 through U4.}
    \label{fig:s_trivial_bulk_modes_2/3}
    \end{adjustwidth*}
\end{figure}

\begin{figure}[htp]
    \begin{adjustwidth*}{-1in}{-1in}
    \hsize=\linewidth
    \includegraphics[width=1.3\textwidth]{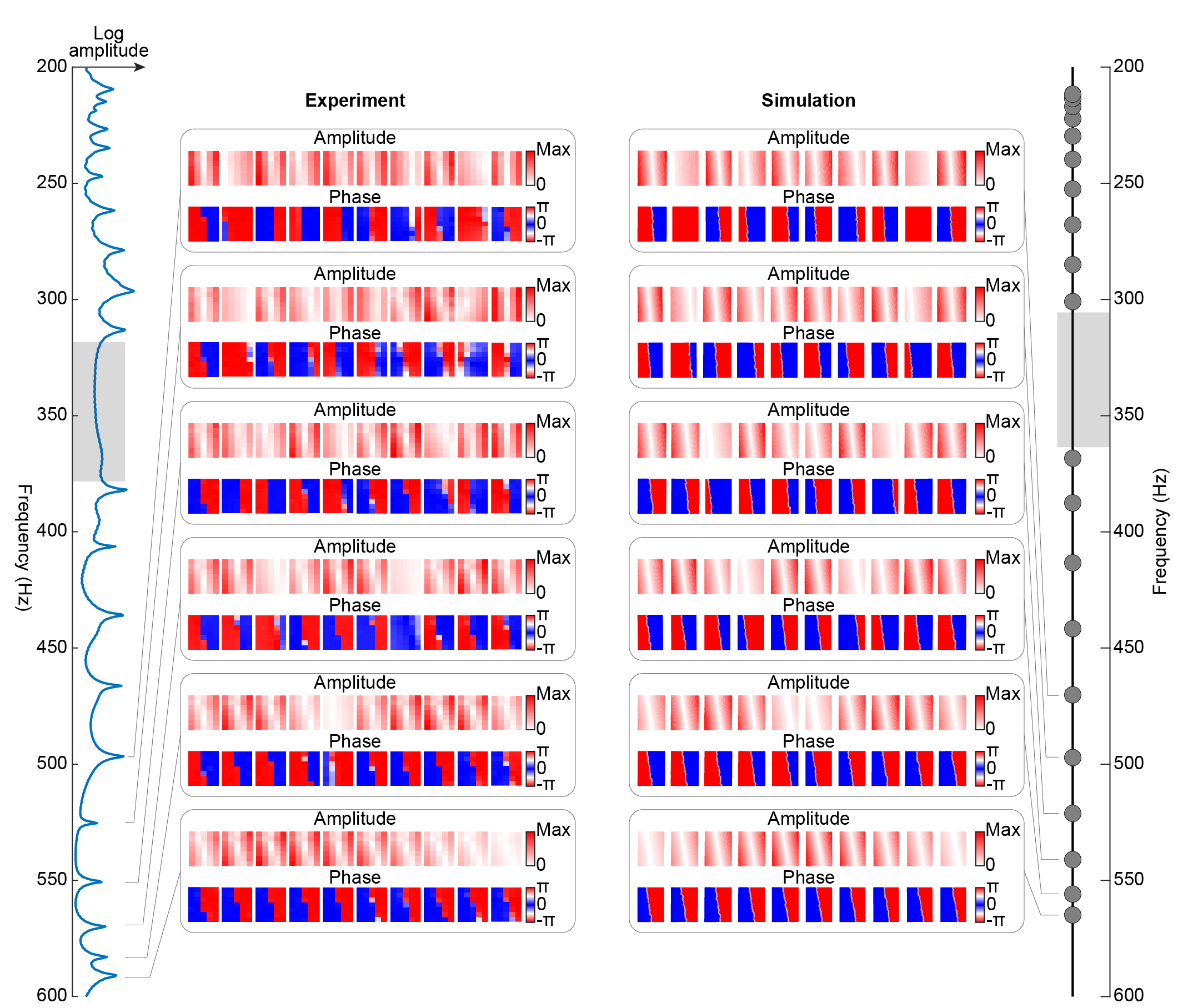}
    \centering
    \caption{
        \textbf{Bulk modes (U5 to U10) for the trivial structure.}
        We show the measured (left half) and the simulated (right half) mode shapes for the bulk mode U5 through U10.}
    \label{fig:s_trivial_bulk_modes_3/3}
    \end{adjustwidth*}
\end{figure}

\FloatBarrier   %

\newpage
\section{Visualization of all modes for the non-trivial structure}
\label{sec:sup_topo}
We show all the bulk modes of the non-trivial Structure with side-to-side comparison to the eigenmode shapes predicted by the FEA software COMSOL. Note that the very flat lower band is too crowded for individual identification of the modes but the agreement between the measured mode shapes and the simulated ones for the upper bulk helps with the identification of the bandgap as well as the two in-gap modes. 

\begin{figure}[htp]
    \begin{adjustwidth*}{-1in}{-1in}
    \hsize=\linewidth
    \includegraphics[width=1.3\textwidth]{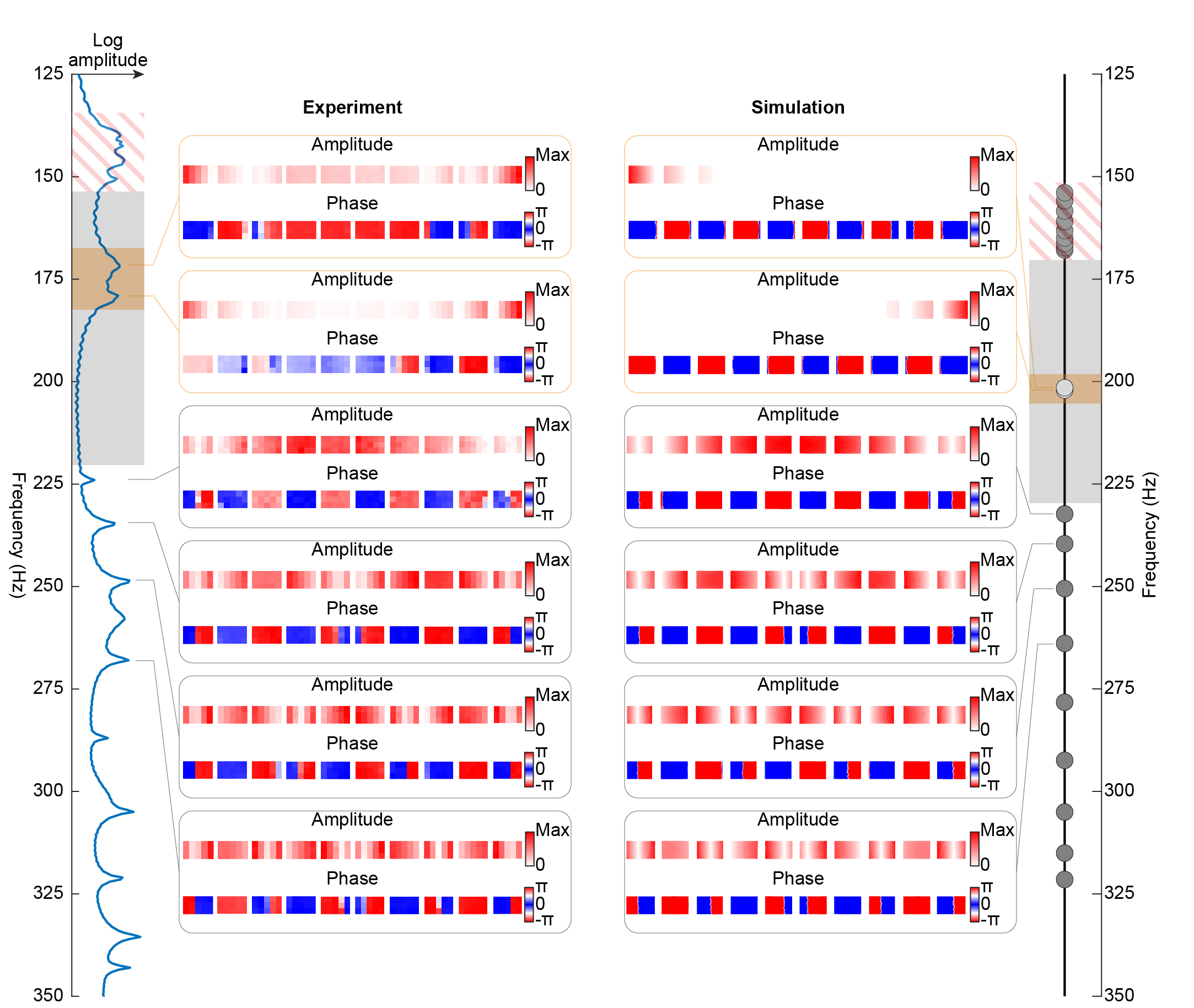}
    \centering
    \caption{
        \textbf{In-gap modes (E1, E2) and some upper bulk modes (U1 to U4) of the non-trivial structure.}
        We show the measured (left half) and the simulated (right half) mode shapes for modes E1 through U4.}
    \label{fig:s_topo_bulk_modes_1/2}
    \end{adjustwidth*}
\end{figure}

\begin{figure}[htp]
    \begin{adjustwidth*}{-1in}{-1in}
    \hsize=\linewidth
    \includegraphics[width=1.3\textwidth]{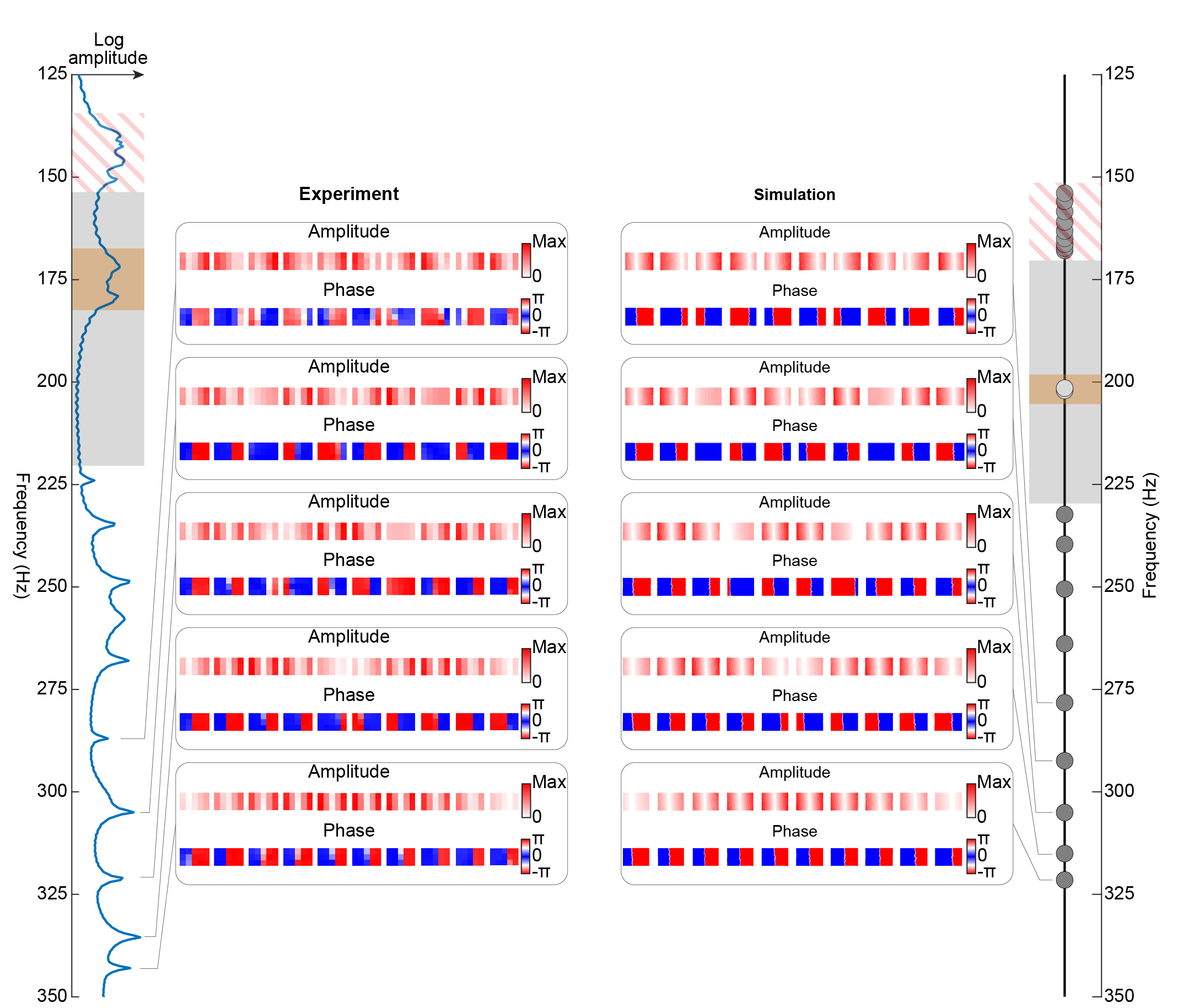}
    \centering
    \caption{
        \textbf{Upper bulk modes (U5 to U9) for the non-trivial structure.}
        We show the measured (left half) and the simulated (right half) mode shapes for the Upper bulk mode U5 through U9.}
    \label{fig:s_topo_bulk_modes_2/2}
    \end{adjustwidth*}
\end{figure}

\FloatBarrier   %

\end{document}